\documentclass[12pt,thmsb]{article}
\usepackage{graphicx}
\usepackage{amsmath}
\usepackage{amsfonts}
\usepackage{amssymb}
\newtheorem{theorem}{Theorem}

\newtheorem{proposition}[theorem]{Proposition}

\newenvironment{proof}[1][Proof]{\textbf{#1.} }{\ \rule{0.5em}{0.5em}}

\begin{document}

\title{Scalar-field cosmologies with an arbitrary potential}
\author{John Miritzis\\Department of Marine Sciences, University of the Aegean\\e-mail: john@env.aegean.gr}
\maketitle
\begin{abstract}
We study the late time evolution of flat and negatively curved FRW models with
a perfect fluid matter source and a scalar field having an arbitrary
non-negative potential function $V\left(  \phi\right)  .$ We prove using a
dynamical systems approach four general results for a large class of
non-negative potentials which show that almost always the universe ever
expands. In particular, for potentials having a local zero minimum, flat and
negatively curved FRW models are ever expanding and the energy density
asymptotically approaches zero. We investigate the conditions under which the
scalar field asymptotically approaches the minimum of the potential. We
discuss the question of whether a closed FRW with ordinary matter can avoid
recollapse due to the presence of a scalar field with a non-negative potential.
\end{abstract}

\section{Introduction}

Scalar fields currently play a prominent role in the construction of
cosmological scenarios aiming to describe the structure and evolution of the
early universe. The standard inflationary idea requires that there be a period
of slow-roll evolution of a scalar field (the inflaton) during which its
potential energy drives the universe in a quasi-exponential expansion
\cite{ko-tu90}. Scalar fields also arise naturally in alternative theories of
gravity which aim to extend general relativity, e.g., higher order gravity
theories, scalar-tensor and string theories. In higher derivative gravity, due
to its conformal relation with general relativity \cite{ba-co88}, scalar
fields appear in the Einstein frame with a self-interaction, nonlinear
potential term which mimics the higher order curvature properties of the
original (Jordan) frame. Generalizations of the original Brans-Dicke theory
lead to scalar-tensor theories with scalar field self-interactions and
dynamical couplings to matter. Further generalizations can be achieved by
considering multiple scalar fields \cite{da-es92}. Similar results can be
proved for simple bosonic string theories \cite{cfmp85}.

Most of the studies of scalar-field cosmologies with the dynamical systems
methods are restricted to Friedmann-Robertson-Walker (FRW) models (see for
example \cite{guetal,cole} and references therein). On the other hand, there
are also important investigations in spatially homogeneous Bianchi cosmologies
with an exponential potential \cite{coibho}, as well as in models containing
both a perfect fluid of ordinary matter and a scalar field with an exponential
potential, the so-called ``scaling'' cosmologies \cite{bicohoibol}.

It is important to investigate the general properties shared by all FRW models
with a scalar field irrespectively of the particular choice of the potential.
In this paper we study the time evolution of flat and negatively curved FRW
models with a perfect fluid and a scalar field with an arbitrary non-negative
potential function $V\left(  \phi\right)  $ and prove some general results for
a large class of such functions. More precisely, we first show that $\phi$
almost always diverges in the past. Furthermore, for potentials having a local
zero minimum, flat and negatively curved FRW models are ever expanding and the
energy density asymptotically approaches zero. If in addition the potential
has a unique global zero minimum, the scalar field almost always
asymptotically approaches zero. Similarly for exponential potentials it is
shown that in an expanding universe the energy density $\rho$ of ordinary
matter and the Hubble function $H$ asymptotically approach zero, while the
scalar field $\phi$ diverges to $+\infty.$

The plan of the paper is as follows. In the next Section we write down the
field equations assuming an arbitrary non-negative potential, as a constrained
four-dimensional dynamical system. In Section 3 and prove the above mentioned
four propositions under mild conditions on the potential. In Section 4 we
discuss the closed universe recollapse conjecture for a $k=1$ FRW universe,
containing a perfect fluid and a scalar field with a non-negative potential.

\section{Scalar-field cosmologies}

In General Relativity the evolution of FRW models with a scalar field
(ordinary matter is described by a perfect fluid with energy density $\rho$
and pressure $p$) are governed by the Friedmann equation,\footnote{We adopt
the metric and curvature conventions of \cite{wael}. Here, $a\left(  t\right)
$ is the scale factor, an overdot denotes differentiation with respect to time
$t,$ and units have been chosen so that $c=1=8\pi G.$}
\begin{equation}
\left(  \frac{\dot{a}}{a}\right)  ^{2}+\frac{k}{a^{2}}=\frac{1}{3}\left(
\rho+\frac{1}{2}\dot{\phi}^{2}+V\left(  \phi\right)  \right)  , \label{fri1jm}%
\end{equation}
the Raychaudhuri equation,
\begin{equation}
\frac{\ddot{a}}{a}=-\frac{1}{6}\left(  \rho+3p+2\dot{\phi}^{2}-2V\right)  ,
\label{fri2jm}%
\end{equation}
the equation of motion of the scalar field,
\begin{equation}
\ddot{\phi}+3\frac{\dot{a}}{a}\dot{\phi}+V^{\prime}\left(  \phi\right)  =0,
\label{emsjm}%
\end{equation}
and the conservation equation,
\begin{equation}
\dot{\rho}+3\left(  \rho+p\right)  \frac{\dot{a}}{a}=0. \label{conssfjm}%
\end{equation}
Here $V\left(  \phi\right)  $ is the potential energy of the scalar field and
$V^{\prime}=dV/d\phi.$ We assume an equation of state of the form
$p=(\gamma-1)\rho,$ with $0\leq\gamma\leq2.$

Formally the FRW model with a scalar field is obtained if we add to the matter
content of the classical Friedmann universe a perfect fluid with energy
density $\varepsilon:=\frac{1}{2}\dot{\phi}^{2}+V$ and pressure $q:=\frac
{1}{2}\dot{\phi}^{2}\!-V$. However, this fluid violates the strong energy
condition, i.e., $\varepsilon+3q=2\dot{\phi}^{2}\!-2V$ may be negative. It is
precisely this violation that leads to inflation in the very early universe
\cite{ko-tu90}.

From Eqs. (\ref{fri1jm})-(\ref{conssfjm}) we see that the state $\left(
a,\dot{a},\rho,\phi,\dot{\phi}\right)  \in\mathbb{R}^{5}$ of the system lies
on the hypersurface defined by the constraint (\ref{fri1jm}) and the remaining
evolution equations can be written as a five-dimensional dynamical system. In
the case of a flat, $k=0$ model, the dimension of the dynamical system reduces
to four.

In the remaining of the paper we assume that the potential of the scalar field
is an arbitrary non-negative (at least $C^{2}$) function. Gradually we
introduce further conditions on the potential that allow us to analyze the
general properties of the resulted dynamical system. Setting $\dot{\phi
}=y,\,\;\dot{a}/a=H$ the evolution equations (\ref{fri2jm})-(\ref{conssfjm})
for the flat model become%
\begin{align}
\dot{\phi}  &  =y,\nonumber\\
\dot{y}  &  =-3Hy-V^{\prime}\left(  \phi\right)  ,\nonumber\\
\dot{\rho}  &  =-3\gamma\rho H,\nonumber\\
\dot{H}  &  =-\frac{1}{2}y^{2}-\frac{\gamma}{2}\rho, \label{sfjm}%
\end{align}
subject to the constraint
\begin{equation}
3H^{2}=\rho+\frac{1}{2}y^{2}+V\left(  \phi\right)  . \label{costrsfjm}%
\end{equation}
Therefore, the phase space of the dynamical system (\ref{sfjm}) is the set
\[
\Sigma:=\left\{  \left(  \phi,y,\rho,H\right)  \in\mathbb{R}^{4}:3H^{2}%
=\rho+\frac{1}{2}y^{2}+V\left(  \phi\right)  \right\}  .
\]

\section{\vspace{0in}Flat and negatively curved FRW with an arbitrary
non-negative potential}

A theorem due to Foster \cite{fost} states that for expanding flat FRW models
in vacuum with a scalar field having a non-negative potential, $\phi$ almost
always diverges in the past. The following Proposition is a generalization of
this result in the case we include a barotropic fluid.

\begin{proposition}
Let $\mathbf{x}\in\Sigma$ and let $O^{-}\left(  \mathbf{x}\right)  $ be the
past orbit of $\mathbf{x}$ under (\ref{sfjm}). Then $\phi$ is almost always
unbounded on $O^{-}\left(  \mathbf{x}\right)  .$
\end{proposition}

\begin{proof}
Let $\mathbf{x}\in\Sigma$ be such that $\phi$ is bounded on $O^{-}\left(
\mathbf{x}\right)  .$ Using the same type of argument as in Theorem 1 of
\cite{fost}, one can show that the $\alpha-$limit set of $\mathbf{x}$ lies on
the plane $\rho=0,$ $y=0.$ From (\ref{sfjm}) we see that the only invariant
sets lying on the plane $\rho=0,$ $y=0,$ are equilibrium points of the form
$\left(  \phi,y,\rho,H\right)  :$ $\rho=0,y=0$ and $\phi_{1}:V^{\prime}\left(
\phi_{1}\right)  =0.$\ Taking into account the constraint (\ref{costrsfjm}),
we see that the past limit set $\alpha\left(  \mathbf{x}\right)  $ consists of
equilibrium points of the form $\left(  \phi_{1},0,0,H_{1}\right)  ,$ where
$H_{1}^{2}=V\left(  \phi_{1}\right)  /3.$

We will show that no such equilibrium is past asymptote to an open
neighborhood of $\mathbf{x}$. To reduce the dimension of the dynamical system
(\ref{sfjm}) we use the constraint (\ref{costrsfjm}) to eliminate $\rho.$ The
evolution equations become%
\begin{align}
\dot{\phi}  &  =y,\nonumber\\
\dot{y}  &  =-3Hy-V^{\prime}\left(  \phi\right)  ,\nonumber\\
\dot{H}  &  =-\frac{3\gamma}{2}H^{2}-\frac{2-\gamma}{4}y^{2}+\frac{\gamma}%
{2}V\left(  \phi\right)  . \label{sys1}%
\end{align}
For an expanding universe, any equilibrium point of (\ref{sys1}) can be
written as $\mathbf{q}=\left(  \phi_{1},0,\sqrt{V_{0}/3}\right)  ,$ where
$\phi_{1}:V^{\prime}\left(  \phi_{1}\right)  =0$ and $V_{0}:=V\left(  \phi
_{1}\right)  >0.$ It is straightforward to see that the eigenvalues of the
Jacobian matrix of (\ref{sys1}) at $\mathbf{q}$ are
\begin{equation}
-\gamma\sqrt{3V_{0}},\qquad\frac{1}{2}\left(  -\sqrt{3V_{0}}\pm\sqrt{\left(
3V_{0}-4V^{\prime\prime}\left(  \phi_{1}\right)  \right)  }\right)  .
\label{eigen}%
\end{equation}
At least two of them have negative real parts and, therefore the center
manifold theorem implies that there exists a 2 or 3-dimensional stable
manifold through $\mathbf{q}$ (see for example \cite{perko}). That means that
all trajectories asymptotically approaching $\mathbf{q}$ as $t\rightarrow
\infty,$ lie on a 2 or 3-dimensional invariant manifold. We conclude that all
solution curves asymptotically approaching $\mathbf{q}$ as $t\rightarrow
-\infty,$ lie on a 0 or 1-dimensional invariant manifold.
\end{proof}

\bigskip

If all of the eigenvalues have negative real parts the assertion follows
immediately from the linearization theorem. In particular, if $V^{\prime
\prime}\left(  \phi_{1}\right)  >0,$ i.e., $\phi_{1}$ is a local minimum of
the potential, then the equilibrium point $\left(  \phi_{1},0,\sqrt{V_{0}%
/3}\right)  $ is future asymptotically stable. This equilibrium solution
corresponds to a vacuum de Sitter space with a cosmological constant equal to
$\sqrt{V_{0}}.$

Unfortunately the above Proposition cannot be applied to potentials having a
zero local minimum $V\left(  \phi_{1}\right)  $, because for $V_{0}=0,$ the
eigenvalues (\ref{eigen}) have zero real parts. However, a large class of
potentials used in scalar-field cosmological models have a zero local minimum.
Examples of potentials belonging to this class are polynomial potentials of
the form $V\left(  \phi\right)  =\lambda\phi^{2n},$ generalized and
logarithmic potentials studied by Barrow and Parsons \cite{paba}, or
potentials in conformally related theories of gravity, for example
\begin{equation}
V\left(  \phi\right)  =V_{\infty}\left(  1-e^{-\sqrt{2/3}\phi}\right)  ^{2}
\label{pote}%
\end{equation}
which arises in the conformal frame of the $R+\alpha R^{2}$ theory
\cite{ba-co88,maed}. To study the late time evolution of these models, we
adopt a different approach.

In the following discussion we assume that the potential function of the
scalar field is non-negative with a local minimum $V\left(  0\right)  =0$, but
otherwise arbitrary. For simplicity, we present only the flat case, $k=0$, but
the results can be extended to the case of negatively curved FRW models (cf.
Remark after Proposition 3).

The equilibrium points of (\ref{sfjm}) are $\left(  \phi,y,\rho,H\right)  :$
$\rho=0,y=0$ and $\phi:V^{\prime}\left(  \phi\right)  =0.$\ Using the
constraint (\ref{costrsfjm}) and the assumed $V_{min}=0$ we see that an
equilibrium point is the $\left(  0,0,0,0\right)  .$ Physically, the
equilibrium point $\left(  0,0,0,0\right)  $, corresponds to an empty
universe. Linearization of (\ref{sfjm}) near the equilibrium point $\left(
0,0,0,0\right)  $ shows that the eigenvalues of the Jacobian matrix at that
point have zero real parts (the assumptions on the potential imply that
$V^{\prime}\left(  \phi\right)  =0$ at $\phi=0$ and also that $V^{^{\prime
\prime}}\left(  0\right)  \geq0$), consequently the Hartman-Grobman Theorem
does not apply \cite{perko}. Therefore, we cannot conclude about the stability
of the equilibrium. However, looking at the system (\ref{sfjm}), one arrives
at the conclusion that for an ever-expanding universe, $H\left(  t\right)  $
and $\rho\left(  t\right)  $ are decreasing functions and $\phi\left(
t\right)  $ oscillates around the minimum of the potential with a decreasing
amplitude due to the damping factor $3H\dot{\phi}.$ These intuitive arguments
are made more precise in Propositions 2 and 3 below.

We first show that an initially expanding flat universe remains
ever-expanding. In fact, the set $\left\{  \left(  \phi,y,\rho,H\right)
\in\mathbb{R}^{4}:H=0\right\}  $ is invariant under the flow of (\ref{sfjm})
with the constraint (\ref{costrsfjm}). Therefore, the sign of $H$ is
invariant. (If the sign of $H$ could change, a solution curve starting with
say, a positive $H$, would pass through the point $\left(  0,0,0,0\right)  $
which is an equilibrium solution of (\ref{sfjm}), thus violating the
fundamental existence and uniqueness theorem of differential equations). We
conclude that if $H\left(  t_{0}\right)  >0,$ then $H\left(  t\right)  >0$ for
all $t>t_{0}.$

In the following Proposition the potential function has a local minimum,
$V\left(  0\right)  =0$, but we do not exclude the existence of other local
minima with a higher value of $V$.

\begin{proposition}
Suppose that $V\geq0$ and $V\left(  \phi\right)  =0\Longleftrightarrow\phi=0$.
Suppose also that if $A\subseteq\mathbb{R}$ is such that $V$ is bounded on
$A$, then $V^{\prime}$ is bounded on $A$. Then $\lim_{t\rightarrow+\infty}%
\rho=0=\lim_{t\rightarrow+\infty}y$.
\end{proposition}

\begin{proof}
Consider the trajectory passing through an arbitrary point $\mathbf{x}%
_{0}=\left(  \phi,y,\rho,H\right)  \in\Sigma$ with $H>0$ at $t=t_{0}.$ Since
$H\left(  t\right)  $ is decreasing and positive, it follows that
$\lim_{t\rightarrow\infty}H\left(  t\right)  $ exists and is a non-negative
number $\eta$; also,$\ H\left(  t\right)  \leq H\left(  t_{0}\right)  $, for
all $t\geq t_{0}$. We then deduce from (\ref{costrsfjm}) that each of the
terms $\rho,\frac{1}{2}y^{2}$ and $V$ is bounded by $3H\left(  t_{0}\right)
^{2}$. Let $A=\left\{  \phi:V\left(  \phi\right)  \leq3H\left(  t_{0}\right)
^{2}\right\}  $. Then the trajectory is such that $\phi$ stays inside $A$.\medskip

From (\ref{sfjm}) we get
\[
-\int_{t_{0}}^{t}\left(  \frac{1}{2}y^{2}+\frac{\gamma}{2}\rho\right)
dt=H\left(  t\right)  -H\left(  t_{0}\right)  ,
\]
and taking the limit as $t\rightarrow\infty,$ we obtain%
\[
\frac{1}{2}\int_{t_{0}}^{+\infty}\left(  y^{2}+\gamma\rho\right)  dt=H\left(
t_{0}\right)  -\eta.
\]
Therefore,
\begin{equation}
\int\nolimits_{t_{0}}^{\infty}\left(  y^{2}\left(  t\right)  +\gamma
\rho\left(  t\right)  \right)  dt<\infty. \label{improper}%
\end{equation}
In general, if $f$ is a non-negative function, the convergence of
$\int\nolimits_{t_{0}}^{\infty}f\left(  t\right)  dt$ does not imply that
$\lim_{t\rightarrow\infty}f\left(  t\right)  =0,$ unless the derivative of $f$
is bounded. It can be shown that $\lim_{t\rightarrow\infty}f\left(  t\right)
=0,$ provided that $f^{\prime}$ is bounded from above.\footnote{If $f>0$,
$\dot{f}<k,$ where $k$ is a positive constant and $\int_{t_{0}}^{+\infty
}fdt<+\infty$ then, $\lim_{t\rightarrow+\infty}f\left(  t\right)  =0$. Here is
the proof: Suppose to the contrary that there exists $\varepsilon>0$ such that
one can find arbitrarily large $t$ such that $f\left(  t\right)  >\varepsilon
$. Choose such a $t$ and then define $s_{0}<t$ so that $t-s_{0}=\frac
{\varepsilon}{k}$. One has%
\begin{align*}
\forall s  &  <t,\int_{s}^{t}\dot{f}\left(  s\right)  ds<\left(  t-s\right)
k\Rightarrow f\left(  s\right)  >f\left(  t\right)  -\left(  t-s\right)
k\Rightarrow\\
\int_{s_{0}}^{t}f\left(  s\right)  ds  &  >\left(  t-s_{0}\right)  f\left(
t\right)  -\frac{1}{2}k\left(  t-s_{0}\right)  ^{2}>\frac{\varepsilon^{2}}%
{2k}.
\end{align*}
The assumption $\int_{t_{0}}^{+\infty}fdt<+\infty$ implies that $\lim
_{s\rightarrow+\infty}\int_{s}^{+\infty}fds=0$. This contradicts the fact that
one can find inside any interval $\left(  s,+\infty\right)  $ a subinterval
$\left(  s_{0},t\right)  $ such that $\int_{s_{0}}^{t}f\left(  s\right)
ds>\frac{\varepsilon^{2}}{2k}$.} In our case,
\[
\frac{d}{dt}\left(  y^{2}+\gamma\rho\right)  =2y\dot{y}+\gamma\dot{\rho
}=-6Hy^{2}-2yV^{\prime}\left(  \phi\right)  -3\gamma^{2}\rho H\leq
-2yV^{\prime}\left(  \phi\right)  .
\]
As we already remarked, $y$ is bounded; also, by our assumption on $V$,
$V^{\prime}\left(  \phi\right)  $ is bounded. We conclude that the derivative
of the function $y^{2}+\gamma\rho$ is bounded from above and therefore,
(\ref{improper}) implies that $\lim_{t\rightarrow\infty}y^{2}\left(  t\right)
=0\;$and$\;\lim_{t\rightarrow\infty}\rho\left(  t\right)  =0.$
\end{proof}

\bigskip

The conditions in Proposition 2 are satisfied by a large class of potential
functions, for example by the generalized and logarithmic potentials studied
in \cite{paba}. In particular, this class includes potentials which do not
approach zero as $\phi\rightarrow\infty$ (cf. equation (\ref{pote})).

\textbf{Remark}. It is easy to conclude that $\lim_{t\rightarrow+\infty}%
\rho=0,$ without any assumption of boundedness of $V^{\prime}$. We only have
to consider the case $\eta>0$ because if $\eta=0$ then (\ref{costrsfjm})
entails that $\lim_{t\rightarrow\infty}\rho=0$. By (\ref{sfjm}b), $\rho$ is
decreasing, thus $\exists\lim_{t\rightarrow+\infty}\rho:=\rho_{0}$. We shall
show that $\rho_{0}=0.$ Suppose that $\rho_{0}>0$. Then from (\ref{sfjm}b)
again we get $\dot{\rho}\leq-3\eta\rho_{0}\gamma.$ Thus
\[
\rho\left(  t\right)  -\rho\left(  t_{0}\right)  =\int_{t_{0}}^{t}\dot{\rho
}\left(  s\right)  ds<-3\eta\rho_{0}\gamma\left(  t-t_{0}\right)  ,
\]
which implies that $\rho$ becomes negative for large $t$, a contradiction.

If we assume in addition that $V^{\prime}\left(  \phi\right)  >0$ for $\phi>0$
and $V^{\prime}\left(  \phi\right)  <0$ for $\phi<0$ (thus, $0$ is the only
local minimum of $V$), then we can conclude that, if the initial value of $H$
is less than$\sqrt{V\left(  \infty\right)  /3}$, then $\lim_{t\rightarrow
+\infty}\phi$ is equal to $0$. More precisely, we have the following result.

\begin{proposition}
Suppose that $V^{\prime}\left(  \phi\right)  >0$ for $\phi>0$ and $V^{\prime
}\left(  \phi\right)  <0$ for $\phi<0$. Then, under the assumptions of
Proposition (2), $\lim_{t\rightarrow+\infty}\phi$ exists and is equal to
$+\infty,$ or $-\infty$ or $0$.
\end{proposition}

\begin{proof}
We know that $\exists\lim_{t\rightarrow\infty}H\left(  t\right)  =\eta$. If
$\eta=0$, then from (\ref{costrsfjm}) we obtain $\lim_{t\rightarrow\infty
}V\left(  \phi\right)  =0$. The assumption on the potential implies that
$\lim_{t\rightarrow\infty}\phi=0$. So suppose that $\eta>0$. From
(\ref{costrsfjm}) we get $\lim_{t\rightarrow+\infty}V\left(  \phi\left(
t\right)  \right)  =3\eta^{2}$. Thus there exists $t^{\prime}$ such that
$V\left(  \phi\right)  >3\eta^{2}/2$ for all $t>t^{\prime}$. It follows that
$\phi$ cannot become zero for some $t>t^{\prime}$ because $V\left(  0\right)
=0$. Thus, $\phi$ has a constant sign for all $t>t^{\prime}$.

Suppose that $\phi>0$ for all $t>t^{\prime}$. Since $V$ is increasing as a
function of $\phi$ on $\left(  0,+\infty\right)  $, we have $\lim
_{t\rightarrow+\infty}V\left(  \phi\left(  t\right)  \right)  =3\eta^{2}%
\leq\lim_{\phi\rightarrow+\infty}V\left(  \phi\right)  $. We consider two cases:

(i) If $\lim_{t\rightarrow+\infty}V\left(  \phi\left(  t\right)  \right)
=\lim_{\phi\rightarrow+\infty}V\left(  \phi\right)  $ then obviously
$\lim_{t\rightarrow+\infty}\phi=+\infty$.

(ii) If $\lim_{t\rightarrow+\infty}V\left(  \phi\left(  t\right)  \right)
<\lim_{\phi\rightarrow+\infty}V\left(  \phi\right)  $ then there exists
$\overline{\phi}\geq0$ such that $\lim_{t\rightarrow+\infty}V\left(
\phi\left(  t\right)  \right)  =V\left(  \overline{\phi}\right)  $. Since $V$
is continuous and strictly increasing, it follows that $\lim_{t\rightarrow
+\infty}\phi=\overline{\phi}$. From (\ref{sfjm}b), taking into account that
$\lim_{t\rightarrow+\infty}y=0$ and $H$ is bounded, we deduce that,
$\lim_{t\rightarrow+\infty}\dot{y}=-V^{\prime}\left(  \overline{\phi}\right)
<0.$ Hence, there exists $t^{\prime\prime}>t^{\prime}$ such that for all
$t>t^{\prime\prime}$, $\dot{y}<-V^{\prime}\left(  \overline{\phi}\right)  /2$.
But this entails immediately that%
\[
y\left(  t\right)  -y\left(  t^{\prime\prime}\right)  =\int_{t^{\prime\prime}%
}^{t}\dot{y}dt<\frac{-V^{\prime}\left(  \overline{\phi}\right)  }{2}\left(
t-t^{\prime\prime}\right)  ,
\]
i.e., $y\left(  t\right)  $ takes arbitrarily large negative values as $t$
increases, which is not possible because $\lim_{t\rightarrow+\infty}y=0$.

Thus, if $\phi>0$ for all $t>t^{\prime}$ then $\lim_{t\rightarrow+\infty}%
\phi=+\infty$. Similarly, the case $\phi<0$ for all $t>t^{\prime}$ leads to
$\lim_{t\rightarrow+\infty}\phi=-\infty$.
\end{proof}

\bigskip

It follows that if initially, $3H^{2}\left(  t_{0}\right)  <\min\{\lim
_{\phi\rightarrow+\infty}V\left(  \phi\right)  ,\lim_{\phi\rightarrow-\infty
}V\left(  \phi\right)  \}$, then $\lim_{t\rightarrow+\infty}H\left(  t\right)
=0$. Indeed, we know that $\lim_{t\rightarrow+\infty}\phi$ is equal to $0$ or
$+\infty$ or $-\infty$. If, say, $\lim_{t\rightarrow+\infty}\phi=+\infty$,
then by (\ref{costrsfjm}), $3\eta^{2}=\lim_{t\rightarrow+\infty}V\left(
\phi\left(  t\right)  \right)  =\lim_{\phi\rightarrow+\infty}V\left(
\phi\right)  >3H^{2}\left(  t_{0}\right)  $. This is impossible, since
$H\left(  t\right)  $ is decreasing and thus $H\left(  t_{0}\right)  \geq\eta
$. Likewise, $\lim_{t\rightarrow+\infty}\phi=-\infty$ leads to a
contradiction. Hence, $\lim_{t\rightarrow+\infty}\phi=0$ and this implies that
$\lim_{t\rightarrow+\infty}V\left(  \phi\left(  t\right)  \right)  =0$ and,
again by (\ref{costrsfjm}), $\lim_{t\rightarrow+\infty}H\left(  t\right)  =0$.

We use as an example the potential (\ref{pote}) to visualize the above result.
This potential has a long and flat plateau. For large values of $\phi,$ the
potential $V$ is almost constant, $V_{\infty}=\lim_{\phi\rightarrow+\infty
}V\left(  \phi\right)  ,$ thus $V$ has the general properties for inflation to
commence. According to Proposition 3, if the initial value of $H$ is less than
$V_{\infty},$ then $\phi,H\rightarrow0$ as $t\rightarrow\infty.$ For initial
values of $H$ greater than $V_{\infty},$ $\phi$ may diverge to $+\infty,$
i.e., the scalar field may reach the flat plateau of the potential.

\textbf{Remark}. The above results can be extended to the case of negatively
curved FRW models, $k=-1.$ Setting $x=1/a$ the evolution equations
(\ref{fri2jm})-(\ref{conssfjm}) become%
\begin{align}
\dot{x}  &  =-Hx\nonumber\\
\dot{\phi}  &  =y,\nonumber\\
\dot{y}  &  =-3Hy-V^{\prime}\left(  \phi\right)  ,\nonumber\\
\dot{\rho}  &  =-3\gamma\rho H,\nonumber\\
\dot{H}  &  =-\frac{1}{2}y^{2}-\frac{\gamma}{2}\rho+kx^{2}, \label{k1}%
\end{align}
subject to the constraint
\begin{equation}
3H^{2}+3kx^{2}=\rho+\frac{1}{2}y^{2}+V\left(  \phi\right)  . \label{constk}%
\end{equation}
With slight modifications we can repeat the same type of arguments to show
that Propositions 2 and 3 still hold and, in addition, $\lim_{t\rightarrow
+\infty}x=0.$

Up to now, we have assumed that the potential is non-negative and has a
minimum. Therefore, the above three Propositions do not apply directly to the
important case of an exponential potential, \textit{vis.} $V\left(
\phi\right)  =V_{0}e^{-\lambda\phi},$ with $V_{0}$ and $\lambda$ positive
constants. It is already known that the state corresponding to $\rho=0,$ $y=0$
and $\phi\rightarrow\infty$ is an equilibrium of the projection of the system
(\ref{sfjm}) on the $\phi-y$ plane. This result has been established using
phase plane analysis (cf. \cite{coliwa}) on the reduced two-dimensional
dynamical system. Using similar arguments as in Propositions 2 and 3, we
complete our analysis by showing the global result that, for expanding flat
models with an exponential potential, $\phi\rightarrow\infty$ as
$t\rightarrow\infty.$ More precisely, we have the following result.

\begin{proposition}
Let $V$ be a potential function with the following properties:\footnote{Note
that $V\left(  \phi\right)  =V_{0}e^{-\lambda\phi}$ has all these properties.
We do not exclude potential functions with $\lim_{\phi\rightarrow+\infty
}V\left(  \phi\right)  >0$.}

1. $V$ is non-negative and $\lim_{\phi\rightarrow-\infty}V\left(  \phi\right)
=+\infty$.

2. $V^{\prime}$ is continuous and $V^{\prime}\left(  \phi\right)  <0$.

3. If $A\subseteq\mathbb{R}$ is such that $V$ is bounded on $A$, then
$V^{\prime}$ is bounded on $A$.

Then $\lim_{t\rightarrow+\infty}y=0=\lim_{t\rightarrow+\infty}\rho,$ and
$\lim_{t\rightarrow+\infty}\phi=+\infty$.
\end{proposition}

\begin{proof}
It is easy to see that the set $\rho>0$ is invariant under the flow of
(\ref{sfjm}), therefore $\rho$ is nonzero if initially $\rho\left(
t_{0}\right)  $ is nonzero. Now from (\ref{conssfjm}) it follows that $H$ is
never zero, thus it cannot change sign. Hence, $H$ is always non-negative if
$H\left(  t_{0}\right)  >0$. Further, $H$ is decreasing in view of
(\ref{sfjm}d), thus $\exists\lim_{t\rightarrow+\infty}H=\eta\geq0$.

We know from (\ref{sfjm}d) that%
\begin{equation}
\frac{1}{2}\int_{t_{0}}^{+\infty}\left(  y^{2}+\gamma\rho\right)  dt=H\left(
t_{0}\right)  -\eta<+\infty. \label{*}%
\end{equation}
As in Proposition 2, $\frac{d}{dt}\left(  y^{2}+\gamma\rho\right)
\leq-2yV^{\prime}\left(  \phi\right)  $. From this and (\ref{*}) we deduce
that $\lim_{t\rightarrow+\infty}y=\lim_{t\rightarrow+\infty}\rho=0$, as in
Proposition 2.

We now show that $\lim_{t\rightarrow+\infty}\phi=+\infty$ as in Proposition 3.
It follows from (\ref{costrsfjm}) that $\lim_{t\rightarrow+\infty}V\left(
\phi\right)  =3\eta^{2}$. Since $V$ is strictly decreasing as a function of
$\phi$, we have $V\left(  \phi\right)  >\lim_{\phi\rightarrow+\infty}V\left(
\phi\right)  $ for all $\phi$, thus $\lim_{t\rightarrow+\infty}V\left(
\phi\left(  t\right)  \right)  \geq\lim_{\phi\rightarrow+\infty}V\left(
\phi\right)  $. We consider two cases:

(i) If $\lim_{t\rightarrow+\infty}V\left(  \phi\left(  t\right)  \right)
=\lim_{\phi\rightarrow+\infty}V\left(  \phi\right)  $ then obviously
$\lim_{t\rightarrow+\infty}\phi=+\infty$.

(ii) If $\lim_{t\rightarrow+\infty}V\left(  \phi\left(  t\right)  \right)
>\lim_{\phi\rightarrow+\infty}V\left(  \phi\right)  $, then there exists a
unique $\overline{\phi}$ such that $\lim_{t\rightarrow+\infty}V\left(
\phi\left(  t\right)  \right)  =V\left(  \overline{\phi}\right)  $. It then
follows that $\lim_{t\rightarrow+\infty}\phi\left(  t\right)  =\overline{\phi
}$. From (\ref{sfjm}b) we deduce that $\lim_{t\rightarrow+\infty}\dot
{y}=-V^{\prime}\left(  \overline{\phi}\right)  >0$; thus, there exists
$t^{\prime}$ such that $\dot{y}\geq-V^{\prime}\left(  \overline{\phi}\right)
/2$ for all $t>t^{\prime}$. From this we get that $y\left(  t\right)
-y\left(  t^{\prime}\right)  >-V^{\prime}\left(  \overline{\phi}\right)
\left(  t-t^{\prime}\right)  /2$, which is not possible because $\lim
_{t\rightarrow+\infty}y=0$. We conclude that $\lim_{t\rightarrow+\infty}%
\phi=+\infty$.
\end{proof}

\bigskip

If in addition the potential is such that $\lim_{\phi\rightarrow+\infty
}V\left(  \phi\right)  =0,$ then we conclude that $H\rightarrow0$ as
$t\rightarrow\infty.$ This is obviously true for an exponential potential.

\section{Discussion and the $k=+1$ FRW model}

In this paper we have investigated the qualitative properties of flat and
negatively curved FRW models containing a barotropic fluid and a scalar field
with an arbitrary non-negative potential. For potentials having a unique
minimum $V\left(  0\right)  =0,$ we have shown that in an expanding universe,
the energy density $\rho$ of ordinary matter, the Hubble function $H$ and the
scalar field $\phi$ asymptotically approach zero. This result was rigorously
proved without referring to the exact details of the potential.

However, the situation is more delicate for positively curved FRW models with
a scalar field. In the present state of the universe the scalar field
oscillates around the minimum of the potential and it is unobservable, in the
sense that the energy density $\rho$ of ordinary matter dominates over the
energy density $\varepsilon$ of the scalar field, hence we expect an almost
classical Friedmannian evolution. Using (\ref{emsjm}) it is easy to see that
the energy density of the field satisfies%
\[
\dot{\varepsilon}=-3H\dot{\phi}^{2}%
\]
i.e., in an expanding universe $\varepsilon$ is a decreasing function of time.
Since the energy density of ordinary matter also decreases, it may happen that
in a future time, $\varepsilon$ be comparable to $\rho.$ In particular, for
closed ($k=1$) models, once the scale factor reaches its maximum value and
recollapse commences i.e., $H<0,$ the term $3H\dot{\phi}$ in (\ref{emsjm}) is
no longer a damping factor, but acts as a driving force which forces the field
$\phi$ to oscillate with larger and larger amplitude. If this be the case, the
repulsive effect of the cosmological term may drastically change the evolution
of a classical FRW model.

These intuitive ideas show that it is of interest to study the late time
evolution of the dynamical system (\ref{k1}) with $k=+1$. The main problem we
are faced with is the following: Can a closed universe filled with ordinary
matter and a scalar field avoid recollapse? In the following we give a partial
answer to this question for an arbitrary non-negative potential having a
unique minimum $V\left(  0\right)  =0$.

A closed Friedmann universe is considered almost synonymous to a recollapsing
universe. This is mainly due to our experience with the dust and radiation
filled Friedmann models usually treated in textbooks. That this picture is
misleading follows clearly from an example found by Barrow \emph{et al}
\cite{bgt} according to which an expanding Friedmann model with spatial
topology $S^{3}$ satisfying the weak, the strong, the dominant energy
conditions and the generic condition may not recollapse. Thus the problem of
recollapse of a closed universe to a second singularity is not trivial already
in the Friedmann case.

Among the known global results concerning the closed universe recollapse
conjecture, let us recall that a closed universe recollapses provided that the
strong energy condition (SEC) is satisfied and there exists a maximal
spacelike hypersurface $\Sigma,$ i.e., the expansion is zero on $\Sigma$ (cf.
\cite{mati}). Therefore, in FRW models it suffices to check if the scale
factor has a maximum.

In the following we assume an equation of state of the form $p=(\gamma
-1)\rho,$ $\gamma\in\left(  \frac{2}{3},2\right)  .$\footnote{The range of
$\gamma$ is chosen in accordance to the conditions for recollapse of Barrow
\emph{et al }\cite{bgt}.} Equation (\ref{conssfjm}) can be integrated to give
\begin{equation}
\rho=\mathrm{const\times}a^{-3\gamma}. \label{den}%
\end{equation}
Assume that at time $t_{0}$ (now) the values of $\phi$ and $\dot{\phi}$ are
very small in the sense that
\begin{equation}
\varepsilon_{0}:=\frac{1}{2}\dot{\phi}_{0}^{2}+V_{0}\ll\rho_{0}\text{\textrm{
\ \ and \ }}2\dot{\phi}_{0}^{2}-2V_{0}\ll\rho_{0}+3p_{0} \label{inco}%
\end{equation}
so that the total stress-energy tensor satisfies the SEC. This is a plausible
assumption since the scalar field is unobservable in the present universe. We
write (\ref{emsjm}) (in first-order approximation) as
\begin{equation}
\ddot{\phi}+3H\dot{\phi}+m^{2}\phi=0 \label{line}%
\end{equation}
where $m^{2}:=V^{\prime\prime}\left(  0\right)  .$ This is the equation of
motion of an harmonic oscillator with a variable damping factor $3H\!.$ For a
slowly varying function $H$ this equation can be solved using the
Kryloff-Bogoliuboff \cite{krbo} approximation. We find that the amplitude of
the scalar field varies as $\sim a^{-3/2}.$ Since the amplitude of $\dot{\phi
}$ has the same time dependence and in our approximation $\varepsilon=\frac
{1}{2}\dot{\phi}^{2}+\frac{1}{2}m^{2}\phi^{2}$, it follows that
\begin{equation}
\varepsilon\sim a^{-3}. \label{ener}%
\end{equation}
Comparing this with the time dependence of the density $\rho\sim a^{-3\gamma}$
we arrive at the following picture for the evolution of these universes. If
$\gamma\leq1$ the initial conditions (\ref{inco}) imply that
\[
\varepsilon=\frac{1}{2}\dot{\phi}^{2}+V\ll\rho\text{\textrm{ \ and\ }}%
2\dot{\phi}^{2}-2V\ll\rho+3p,
\]
for all $t\geq t_{0},$ that is, the SEC on the total stress-energy tensor is
satisfied for all $t\geq t_{0}.$ Hence the universe follows the classical
Friedmannian evolution and has a time of maximum expansion. Therefore, with
initial conditions (\ref{inco}) the approximation (\ref{line}) suggests that
if $\gamma\leq1,$ the universe recollapses. The case $\gamma=1$ is
particularly interesting since it corresponds to a dust-filled universe which
perfectly approximates our Universe. For $\gamma>1$, since $\rho$ decreases
faster than the energy density $\varepsilon$ of the scalar field,
$\varepsilon$ eventually dominates over $\rho.$ Hence this model may, or may
not have a time of maximum expansion.

The above analysis may serve as an indication of the richness of possible
scenarios for expanding closed FRW models with a scalar field. However,
spurious conclusions may be drawn from an approximate analysis, especially if
it is referred to the asymptotic behavior of the solutions. Similarly, the
determination of the stability of equilibrium solutions does not reveal the
complete global behavior of a dynamical system of dimension greater than two.
A rigorous proof of the closed universe recollapse conjecture may come from
the investigation of the global structure of the solutions of (\ref{k1}) with
$k=+1$.

\bigskip

\textbf{Acknowledgments} I thank N. Hadjisavvas and S. Cotsakis for fruitful
discussions during the preparation of this work.

\bigskip
\end{document}